\documentclass[aip,prl,reprint]{revtex4-2} 
\usepackage[T1]{fontenc}
\usepackage{times}

\usepackage{graphicx}
\usepackage{sidecap}
\usepackage{booktabs}

\usepackage{amsfonts, amsmath,amssymb, bm, graphicx}
\usepackage{siunitx}

\usepackage{amssymb}
\usepackage[english]{babel}

\newcommand{\figref}[2]{\hyperref[#1]{\ref{#1}(#2)}}
\newcommand{\figrefsub}[3]{\hyperref[#1]{\ref{#1}(#2)#3}}

\usepackage{physics}
\usepackage{lipsum}
\usepackage{changes}

\usepackage[colorlinks=true,allcolors=blue]{hyperref}
\usepackage{footmisc}

\usepackage{upgreek}

\usepackage{soul}

\begin{document}

\title{Control of magnon frequency combs in magnetic rings} 

\author{Christopher Heins}
\affiliation{Helmholtz-Zentrum Dresden--Rossendorf, Institut f\"ur Ionenstrahlphysik und Materialforschung, D-01328 Dresden, Germany}
\affiliation{Fakult\"at Physik, Technische Universit\"at Dresden, D-01062 Dresden, Germany}

\author{Attila K\'akay}
\affiliation{Helmholtz-Zentrum Dresden--Rossendorf, Institut f\"ur Ionenstrahlphysik und Materialforschung, D-01328 Dresden, Germany}

\author{Joo-Von Kim}
\affiliation{Centre de Nanosciences et de Nanotechnologies, CNRS, Universit\'e Paris-Saclay, 91120 Palaiseau, France}

\author{Gregor Hlawacek}
\affiliation{Helmholtz-Zentrum Dresden--Rossendorf, Institut f\"ur Ionenstrahlphysik und Materialforschung, D-01328 Dresden, Germany}

\author{J\"urgen Fassbender}
\affiliation{Helmholtz-Zentrum Dresden--Rossendorf, Institut f\"ur Ionenstrahlphysik und Materialforschung, D-01328 Dresden, Germany}
\affiliation{Fakult\"at Physik, Technische Universit\"at Dresden, D-01062 Dresden, Germany}

\author{Katrin Schultheiss}\email{k.schultheiss@hzdr.de}
\affiliation{Helmholtz-Zentrum Dresden--Rossendorf, Institut f\"ur Ionenstrahlphysik und Materialforschung, D-01328 Dresden, Germany}

\author{Helmut Schultheiss}\email{h.schultheiss@hzdr.de}
\affiliation{Helmholtz-Zentrum Dresden--Rossendorf, Institut f\"ur Ionenstrahlphysik und Materialforschung, D-01328 Dresden, Germany}

\date{\today}


\begin{abstract}
    
Using Brillouin light scattering microscopy, we study the rich dynamics in magnetic disks and rings governed by nonlinear interactions, focusing on the role of vortex core dynamics on the spin-wave eigenmode spectrum. By strongly exciting quantized magnon modes in magnetic vortices, self-induced magnon Floquet states are populated by the intrinsic nonlinear coupling of magnon modes to the vortex core gyration. In magnetic rings, however, this generation is suppressed even when exciting the system over a large power range. To retrieve the rich nonlinear dynamics in rings, we apply external in-plane magnetic fields by which the vortex core is restored. Our findings demonstrate how to take active control of the nonlinear processes in magnetic structures of different topology.

\end{abstract}

\maketitle


Nonlinear interactions in magnetic systems have potential for advancing unconventional computing technologies. Magnonic systems exhibit complex and tunable dynamics driven by nonlinear interactions which enable a variety of behaviors, including mode coupling and frequency conversion. These phenomena can be used for unconventional computing\cite{papp_nanoscale_2021,finocchio_roadmap_2024}, particularly in the context of reservoir computing \cite{watt_reservoir_2020,nakane_performance_2023,namiki_fast_2024, nagase_spin-wave_2024}. Magnetic vortices offer a unique platform for studying these interactions due to their stability and topology.
By manipulating the dynamics of magnetic vortices and controlling magnon scattering processes, it becomes possible to perform pattern recognition tasks on time-series data\cite{korber_pattern_2023}. 
Exploring the underlying mechanisms of nonlinear magnon scattering in magnetic vortices, particularly involving the core gyration, could, therefore, open up new avenues for innovative computing technologies.
In this work, we investigate the fundamental mechanisms of nonlinear magnon scattering in magnetic disks and rings and explore how these interactions can be controlled actively.

\begin{figure}
    \includegraphics{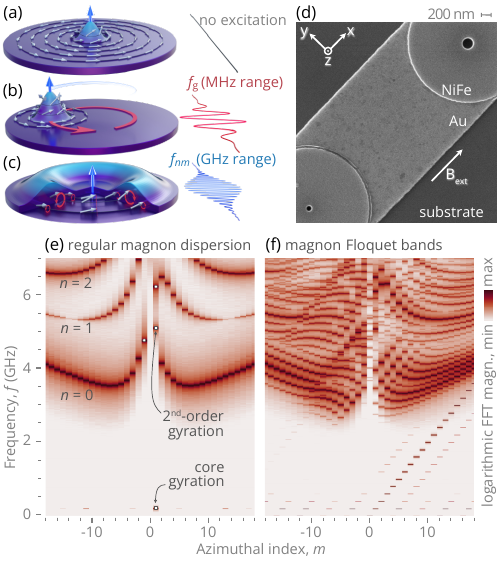}
     \caption{(a) Schematic illustrations of (a) a magnetic vortex, (b) the vortex core gyration, and (c) the fundamental magnon mode ($0,0$) with the vortex core remaining quasi-static in the center of the disk. Dimensions are not drawn to scale. (d) Helium-ion microscope image of the studied sample. (e),(f) Simulated dispersion relations of (e) the regular magnon modes in a vortex with the core static in its center and (f) magnon Floquet bands in a vortex with its core gyrating. }
    \label{fig:schematic}
\end{figure}

\begin{SCfigure*}
    \includegraphics{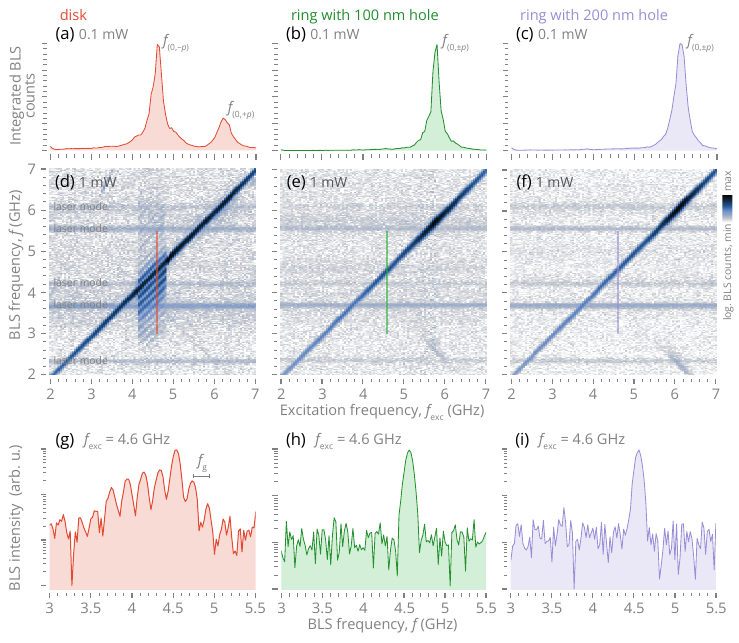}
    \caption[c]{BLS spectra of the magnon modes measured in the linear regime on (a) a \SI{2}{\micro\meter} diameter disk, (b) a ring with an outer diameter of \SI{2}{\micro\meter} and a \SI{100}{\nano\meter} diameter central hole, and (c) a ring with an outer diameter of \SI{2}{\micro\meter} and a \SI{200}{\nano\meter} wide central hole. (d) BLS spectra measured as a function of excitation frequency for a higher power reaching the nonlinear regime. In the disk, this leads to the generation of magnon frequency combs. (e),(f) Removing the vortex core completely suppresses the generation of Floquet side bands and, thereby, the frequency comb. (g)-(i) BLS spectra extracted for an excitation frequency of $f_\mathrm{exc}=\SI{4.6}{\giga\hertz}$. (g) The magnon Floquet modes measured in the disk are equally spaced by the frequency of the gyrotropic mode $f_\mathrm{g}=\SI{205}{\mega\hertz}$. }
    \label{fig:rf-sweep}
    \end{SCfigure*}

In ferromagnetic disks with tens of nanometers in thickness and a few micrometers in diameter, the ground state is a magnetic vortex\cite{cowburnSingleDomainCircularNanomagnets1999,shinjo_magnetic_2000,scholz_transition_2003}. As depicted in Fig.~\figref{fig:schematic}{a}, it is characterized by the magnetic moments (white arrows) curling in-plane around the vortex core (blue arrow). In this narrow region in the center of the disk, the magnetization is aligned perpendicularly to the plane with a well-defined polarity $p$, either pointing up ($p=1$) or down ($p=-1$). 
Such magnetic vortices host different dynamic excitations, the lowest order of which is the vortex core gyration\cite{novosad_magnetic_2005, guslienko_vortex-state_2005, stoll_imaging_2015}. This low amplitude translational motion of the vortex core around its equilibrium position in the center of the disk is sketched in Fig.~\figref{fig:schematic}{b}. For the disk dimension studied in this work, the frequency $f_\mathrm{g}$ of the fundamental gyrotropic mode is in the range of a few hundred MHz\cite{guslienko_eigenfrequencies_2002, park_interactions_2005}. Additionally, higher frequency magnon modes also exist around the static vortex core. They can be classified by their radial and azimuthal mode indices ($n,m$), with $n,m$ counting the number of nodes along the disk radius and in angular direction, respectively\cite{guslienko_vortex-state_2005,vogt_optical_2011, schultheissExcitationWhisperingGallery2019, korberNonlocalStimulationThreeMagnon2020}. Figure~\figref{fig:schematic}{c} depicts the lowest-order magnon mode ($0,0$) with its frequency $f_{nm}$ in the GHz range. 

In previous work\cite{heins_self-induced_2024}, we demonstrated that the magnon eigenmode spectrum changes significantly if the vortex core gyration is excited simultaneously with the magnon modes. Figure~\figref{fig:schematic}{e} shows the simulated magnon dispersion for a \SI{50}{\nano\meter} thick, \SI{2}{\micro\meter} diameter Ni$_{81}$Fe$_{19}$ disk with the core static in the center of the disk. The dispersion relation for different radial mode numbers $n=0,1,2,...$ are well separated. 
In contrast, if the core gyration is excited actively, the translational motion of the core results in a periodic modulation of the ground state which leads to the appearance of magnon Floquet bands, as can be seen in Fig.~\figref{fig:schematic}{f}.
The coupling of the Floquet modes to the vortex gyration results in the formation of magnon frequency combs with the spacing equal to the gyration frequency. 
Moreover, if only high-frequency magnon modes are excited above a certain threshold power, the intrinsic nonlinearity of the magnetic system results in the spontaneous coupling of the magnon modes to the vortex gyration itself. This leads to the generation of self-induced magnon Floquet bands which again manifests in the formation of magnon frequency combs. Here, we demonstrate how to take active control over this process in magnetic ring structures. 

\begin{SCfigure*}
    \includegraphics{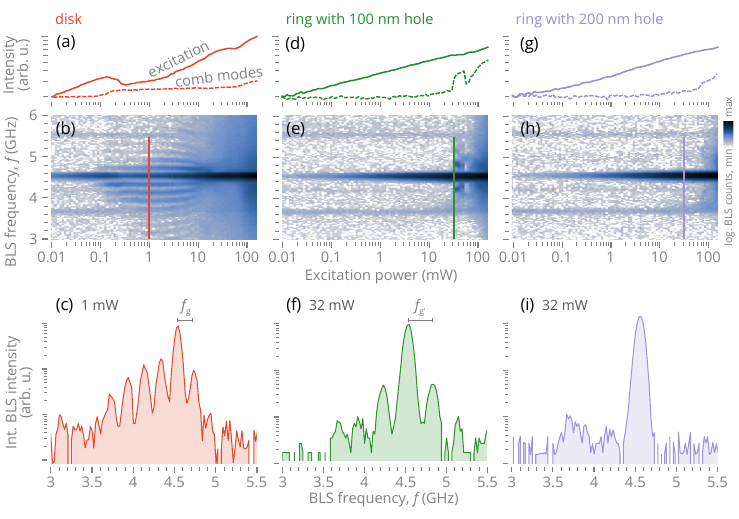}
    \caption{(a,d,g) BLS intensity integrated for the direct excitation frequency $f_\text{exc}=\SI{4.6}{\giga\hertz}$ (solid lines) and the comb modes (dashed lines) measured as a function of the excitation power for the different geometries. (b,e,h) BLS spectra measured as a function of the excitation power. (c,f,i) BLS spectra extracted for (c) \SI{1}{\milli\watt} excitation power in the disk, (f) \SI{32}{\milli\watt} in the ring with the \SI{100}{\nano\meter} hole, and (i) \SI{32}{\milli\watt} in the ring with the \SI{100}{\nano\meter} hole. }
    \label{fig:power-sweep}
    \end{SCfigure*}

For our studies, we pattern a Cr($\SI{5}{\nano\meter}$)/Au($\SI{65}{\nano\meter}$) layer to a coplanar waveguide (CPW) with a \SI{2}{\micro\meter} wide signal line using electron beam lithography, electron beam evaporation and subsequent lift-off [Fig.~\figref{fig:schematic}{d}]. The gap to the \SI{5}{\micro\meter} wide ground lines is \SI{3}{\micro\meter}.
In a second step, we fabricate \SI{50}{\nano\meter} thick  Ni$_{81}$Fe$_{19}$ disks with \SI{2}{\micro\meter} in diameter directly on top of the CPW's signal line, again using electron beam lithography, electron beam evaporation and lift-off. Applying microwave currents to the CPW allows for the direct excitation of magnetization dynamics with a well-defined frequency $f_\mathrm{exc}$. However, due to its symmetry, the in-plane microwave field only couples to the vortex gyration and first-order azimuthal modes with $m\pm 1$.

To study the influence of topology of the magnetic texture on the formation of magnon frequency combs, we remove the vortex core from selected structures. Therefore, we cut well-defined holes in the center of two disks with a helium-ion microscope (HIM)\cite{hlawacekHeliumIonMicroscopy2014,hoflichRoadmapFocusedIon2023} using a \SI{15}{\kilo\electronvolt}, \SI{0.5}{\pico\ampere} Neon beam with a fluence of \SI{1}{\nano\coulomb/\micro \meter^2}, and spot control 6.
One hole is \SI{100}{\nano\meter} in diameter, the other \SI{200}{\nano\meter}, as shown in the HIM image in Fig.~\figref{fig:schematic}{d}. One disk remains unchanged and serves as a reference (not shown in the HIM image). We note that all magnetic structures share one common antenna which allows for the direct comparison of excitation efficiencies across the three structures.

For our experimental studies, we use Brillouin light scattering (BLS) microscopy \cite{sebastian_micro-focused_2015}. 
This technique relies on focusing a monochromatic (\SI{532}{\nano\meter}) continuous-wave laser onto the sample surface using a microscope lens with a high numerical aperture (0.75), which yields a spatial resolution of about \SI{300}{\nano\meter}. The backscattered light is then directed into a Tandem Fabry-P\'{e}rot interferometer\cite{mock_construction_1987} to analyze the frequency shift caused by the inelastic scattering of photons and magnons. The detected intensity of the frequency-shifted signal is directly proportional to the magnon intensity at the respective focusing position. 
During all experiments, the investigated microstructure is imaged using a red LED and a CCD camera. Displacements and drifts of the sample are tracked by an image recognition algorithm and compensated by the positioning system. In order to ensure that magnon modes with different spatial distributions are measured, the signal is integrated over three radial times four azimuthal positions across half the structures. An electromagnet can provide an external magnetic field $B_\text{ext}$ along the microwave antenna. All measurements are performed at room temperature.

We begin characterizing the eigenmodes excited in the linear regime, i.e., at a power of \SI{0.1}{\milli\watt}, just below the threshold for the generation of self-induced magnon Floquet states and magnon frequency combs. 
The magnon response to the microwave excitation  is measured in the disk and rings with \SI{100}{\nano\meter} and \SI{200}{\nano\meter} diameter holes and then integrated in a \SI{200}{\mega\hertz} frequency window around the excitation frequency, respectively. The results are plotted in Fig.~\figref{fig:rf-sweep}{a-c} for excited modes in the frequency range of  \SIrange[]{2}{7}{\giga\hertz}. In the disk [Fig.~\figref{fig:rf-sweep}{a}], the hybridization of the first order azimuthal modes $(0,-1)$ and $(0,+1)$ with the vortex core lifts their degeneracy, resulting in the detection of two peaks at frequencies $f_{0,-p} <  f_{0,+p}$ with a frequency splitting of \SI{1.6}{\giga\hertz}, similar to values reported before \cite{park_interactions_2005,guslienkoDynamicOriginAzimuthal2008}. Which of the two modes has the lower frequency depends on the core polarity $p$ (with $|p|=1$) and is inaccessible in our experiments. In contrast, the rings without the vortex core only show one peak that corresponds to the degenerate modes at frequency $f_{0,1} = f_{0,-1}$ [Fig.~\figref{fig:rf-sweep}{b,c}]. The frequencies differ slightly in the two rings because of varying quantization in radial direction. 




Increasing the excitation power to 
\SI{1}{\milli\watt}, we pump the system strong enough to leave the linear regime. In the vortex-state disk, the nonlinear coupling of the magnon modes to the vortex core gyration leads to the formation of self-induced magnon Floquet bands, which appear as frequency combs as shown in Fig.~\figref{fig:rf-sweep}{d}. In this graph, we plot BLS spectra as a function of the excitation frequency with the detected intensity color-coded on a logarithmic scale. For most excitation frequencies, we detect a direct response which corresponds to the intense diagonal line in Fig.~\figref{fig:rf-sweep}{d}. However, in the frequency range between \SI{4.1}{\giga\hertz} and \SI{4.8}{\giga\hertz}, i.e., when the azimuthal mode with the lower frequency $f_{0,-p}$ is excited, the system's response changes drastically and frequency combs with a well-defined frequency spacing are measured. Note that the faint horizontal lines in the spectra correspond to higher-order laser modes.

\begin{SCfigure*}
    \includegraphics{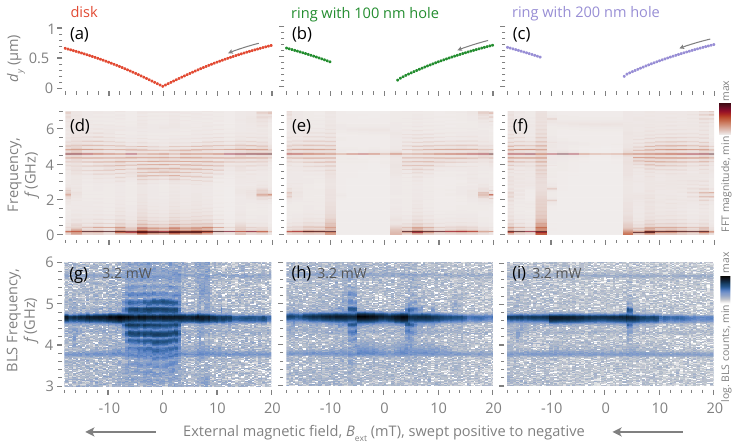}
    \caption{ (a-c) Displacement $d_\text{y}$ of the vortex core from the disk center extracted from micromagnetic simulations for the different geometries. (d-f) Micromagnetic simulations showing the formation of frequency combs in the different geometries when exciting at $f_\mathrm{exc}=\SI{4.6}{\giga\hertz}$. (g-i) BLS spectra measured as a function of a static in-plane magnetic field for the different geometries. The field was swept from positive to negative values. The excitation frequency and power were fixed at $f_\mathrm{exc}=\SI{4.6}{\giga\hertz}$ and \SI{3.2}{\milli\watt}, respectively. }
    \label{fig:field-sweep}
    \end{SCfigure*}

If we extract the frequency for one single excitation frequency, as is shown in Fig.~\figref{fig:rf-sweep}{g} for  $f_\mathrm{exc}=\SI{4.6}{\giga\hertz}$, one can see that the teeth of the frequency comb are equally spaced by $f_\mathrm{g}=\SI{205}{\mega\hertz}$. This matches the frequency of the gyrotropic motion in our system which we determine using micromagnetic simulations with the open-source finite-difference micromagnetics code \textsc{MuMax3}~\cite{vansteenkiste_design_2014}. We model our \SI{50}{\nano\meter} thick, \SI{2}{\micro\meter} diameter disk using $512 \times 512 \times 8$ finite difference cells with $\gamma=\SI{1.86e-11}{\radian/(Ts)}$, $M_\mathrm{s}=\SI{775}{\kilo\ampere/\meter}$, an exchange constant of $A_\mathrm{ex}=\SI{12}{\pico\joule/\meter}$, and $\alpha = 0.007$ –– the nominal value for this material.

Remarkably, self-induced frequency combs are only measured when pumping the lower-frequency azimuthal mode ($0,-p$). We attribute this to the initial spontaneous nonlinear scattering process that couples the magnon modes to the gyrotropic mode and is only possible for ($0,-p$). In the regular magnon dispersion [Fig.~\figref{fig:schematic}{f}], the core polarity is positive ($p=1$) and the modes ($0,-1$), ($0,1$) are indicated by white dots. For any magnon mode to couple to the core gyration, both the frequency and azimuthal mode number need to decrease: $\Delta f=-f_\text{g}$ and $\Delta m = -p$. While for mode ($0,-1$) this is possible via scattering into mode ($0,-2$), mode ($0,1$) does not have a suitable scattering partner. Hence, the spontaneous coupling of this magnon mode to the vortex gyration itself is prohibited and no self-induced magnon Floquet states occur. 

In the next step, we repeat the same measurement on the rings with the \SI{100}{\nano\meter} and \SI{200}{\nano\meter} diameter holes and plot the results in Fig.~\figref{fig:rf-sweep}{e,f}, respectively. In these structures, only the diagonal line of the direct response and no magnon frequency combs are measured over the entire frequency range. The BLS spectra which we extracted for the same excitation frequency of $f_\mathrm{exc}=\SI{4.6}{\giga\hertz}$ [Fig.~\figref{fig:rf-sweep}{h,i}] show only one individual peak. 

These measurements demonstrate how the formation of self-induced magnon Floquet bands in magnetic disks is directly linked to the presence of the vortex core and its gyrotropic motion. If the core is absent, the ground state with circulating in-plane magnetization remains largely unchanged, no self-induced magnon Floquet states emerge and no frequency comb can be measured. 
In the following, we will demonstrate how Floquet bands can be restored even inside magnetic rings by significantly increasing the excitation power or, more efficiently, by applying a small in-plane magnetic field.   

Up to this point, we have discussed the response to the excitation powers at a moderate level of \SI{1}{\milli\watt}. 
In Fig.~\ref{fig:power-sweep}, we present measured BLS spectra for a fixed excitation frequency of $f_\mathrm{exc}=\SI{4.6}{\giga\hertz}$ as a function of the excitation power spanning four orders of magnitude. The frequency is selected from the range for which we measure the frequency comb in Fig.~\figref{fig:rf-sweep}{d}. 
In the disk [Fig.~\figref{fig:power-sweep}{a-c}], powers larger than
\SI{0.1}{\milli\watt} are necessary to excite the magnon frequency comb. Then, the spontaneous nonlinear coupling of the magnon mode to the gyrotropic motion is strong enough to generate self-induced magnon Floquet bands [Fig.~\figref{fig:power-sweep}{b}] and, thereby, a magnon frequency comb. The intensity of the directly excited mode decreases as soon as the Floquet side bands are generated [Fig.~\figref{fig:power-sweep}{a}], indicating the redistribution of energy via nonlinear scattering. 
Frequency combs persist up to powers of
\SI{10}{\milli\watt}. At even stronger excitation, the comb vanishes. Instead, a broad and noisy background signal is measured which we attribute to chaotic switching of the vortex core (confirmed by micromagnetic simulations) and, possibly, to higher-order nonlinear processes. 

In the rings, however, the response to increasing driving powers is quite different. Interestingly, the ring with the \SI{100}{\nano\meter} hole [Fig.~\figref{fig:power-sweep}{d-f}]  shows a frequency comb as well but only at much larger excitation powers above
\SI{30}{\milli\watt} and only inside a small power range. In contrast, no frequency comb is measured at all in the ring with the \SI{200}{\nano\meter} hole [Fig.~\figref{fig:power-sweep}{g-i}]. We attribute this difference to the fact that in the ring with the \SI{100}{\nano\meter} hole, the vortex core can be nucleated at large enough excitation powers to gyrate around the central hole. The \SI{200}{\nano\meter} hole, however, is too large and no vortex core can be nucleated even at high excitation powers. 

As can be seen in the individually extracted BLS spectrum in Fig.~\figref{fig:power-sweep}{c}, the distance between neighboring frequency peaks measured in the disk remains constant around $f_\mathrm{g}=\SI{205}{\mega\hertz}$ and does not change and matches the value discussed above over a large range of excitation powers. 
The spacing of the frequency comb in the ring with the \SI{100}{\nano\meter} hole [Fig.~\figref{fig:power-sweep}{f}], however, is larger ($f_\mathrm{g'}=\SI{303}{\mega\hertz}$) which we attribute to its gyration around the hole. 

Since the formation of self-induced magnon Floquet states is directly linked to the presence of the vortex core, static in-plane magnetic fields provide another means to control its existence. When applying a small in-plane magnetic field, the vortex core inside a magnetic disk shifts perpendicularly to the field direction to minimize the Zeeman energy. To demonstrate this, we perform micromagnetic simulations using \textsc{MuMax3}~\cite{vansteenkiste_design_2014}. 
We start with a vortex state at \SI{0}{\milli\tesla}, then first increase the static in-plane field $B_\text{ext}$ in \SI{0.5}{\milli\tesla}  steps along the $x$-direction up to \SI{20}{\milli\tesla} and, finally, decrease the field with the same step size to \SI{-18}{\milli\tesla}. At each field, the ground state is found by minimizing the energy. In order to promote the nucleation of the core, a small additional field $B_\text{bias}=\SI{0.1}{\milli\tesla}$ is applied along the $z$-direction. From this simulation data, we extract the $y$-position of the vortex core and plot its displacement $d_{y}$ from the disk center as a function of the field in Fig.~\figref{fig:field-sweep}{a-c}. 

In the disk, the core moves gradually with the applied field and its position is symmetric around zero field. In contrast, this is not the case anymore in the rings where the core vanishes for a certain field range. Coming from large fields, the core is stable up to \SI{2.5}{\milli\tesla} (\SI{3.5}{\milli\tesla}) for the ring with the \SI{100}{\nano\meter} (\SI{200}{\nano\meter}) hole, respectively.  Increasing the field in the negative direction, a threshold needs to be overcome to nucleate the vortex core which is reached at \SI{-10}{\milli\tesla} (\SI{-12}{\milli\tesla}) for the smaller (larger) hole. This asymmetry reverses if the magnetic field is swept in the opposite direction from negative to positive fields leading to a hysteric behavior. The smaller the hole, the more narrow the field range without vortex core.

Next, we simulate the dynamics in the disks under the influence of the same static in-plane magnetic fields. We start analogous to the static simulation with a vortex state at \SI{0}{\milli\tesla}. Then the static in-plane field $B_\text{ext}$ is increased in \SI{1}{\milli\tesla} steps and at each field the ground state is found by minimizing the energy. When reaching a field of \SI{20}{\milli\tesla}, the dynamics are excited by a dynamic field in $y$-direction $B_y=\SI{0.5}{\milli\tesla}\cdot\sin{(2\pi\cdot\SI{4.6}{\giga\hertz}\cdot t)}$ for \SI{26}{\nano\second}. The static field is then decreased in \SI{2}{\milli\tesla} steps and at each field the magnetization dynamics are recorded. For each field step we calculate the power spectrum as plotted for the different geometries in Fig.~\figref{fig:field-sweep}{d-f}. 

From these dynamic simulations, it is evident that frequency combs are generated whenever a core is present in the structure. Only at the positive field limit, when the core vanishes, there is slight mismatch between the presence of the core in the static and dynamic simulations. In the dynamic simulations, the core vanishes at slightly larger fields already. This is attributed to the orbit of the gyrotropic motion which forces the core into the holes while it is still stable under static conditions. 

To confirm these micromagnetic simulations, we measure BLS spectra on the different structures as a function of the externally applied magnetic field $B_\text{ext}$ and plot the results in Fig.~\figref{fig:field-sweep}{g-i}. In these measurements, the field is swept gradually from its maximum value of \SI{+50}{\milli\tesla} to \SI{-30}{\milli\tesla}. The graphs focus on the same field range as the micromagnetic simulations. The excitation frequency and power were fixed at $f_\mathrm{exc}=\SI{4.6}{\giga\hertz}$ and \SI{3.2}{\milli\watt}, respectively, corresponding to the condition for efficient comb generation without external magnetic field [Fig.~\figref{fig:power-sweep}{a}]. Note that these conditions change slightly when an external magnetic field is applied.

In contrast to the simulations, frequency combs are only observed in the disk for a limited field range $\SI{3.5}{\milli\tesla}>B_\text{ext}> \SI{-7}{\milli\tesla}$ [Fig.~\figref{fig:field-sweep}{g}]. For positive fields, this is related to the different initial conditions in experiment and simulation. While the core is never expelled from the disk in simulation, the maximum applied field \SI{+50}{\milli\tesla} is strong enough to saturate the disk in the experiments. Therefore, when reducing the field, two vortices nucleate at the disk boundary which merge to a single vortex at small magnetic fields only\cite{korber_modification_2023}. Furthermore, at larger negative fields, the vortex core is expelled at much lower field magnitudes in experiments compared to micromagnetic simulations \cite{korber_modification_2023}. In both cases without the vortex core, no self-induced magnon Floquet states can be generated anymore and only the direct excitation is measured. 

In the ring with the \SI{100}{\nano\meter} diameter hole [Fig.~\figref{fig:field-sweep}{h}], a frequency comb appears at about \SI{6}{\milli\tesla} which is again related to saturating the structure when starting the measurement. However, contrary to the disk, the frequency comb disappears again at \SI{4}{\milli\tesla} already, only to reappear in a small field range $\SI{-5}{\milli\tesla}>B_\text{ext}> \SI{-6.5}{\milli\tesla}$. 
In the ring with the larger hole [Fig.~\figref{fig:field-sweep}{i}], a faint frequency comb is measured for positive fields only ($\SI{6.5}{\milli\tesla}>B_\text{ext}> \SI{4}{\milli\tesla}$). The disappearance of frequency combs around zero magnetic field is consistent with the core vanishing inside the holes.

In conclusion, we have demonstrated that the presence of magnon Floquet states is strongly linked to the presence of the vortex core. If the core is removed from the spin texture, even strong modulations of the ground state with low frequencies will not create magnon frequency combs. The reappearance of the magnon frequency comb at very high powers and zero static magnetic field is linked to the creation of a new vortex. This in itself is already an intriguing observation: Strong spin-wave excitation can create a vortex core without the need of a bias magnetic field. We showed, that this power threshold for the creation of a vortex core via spin waves can be significantly lowered by applying a small static in-plane magnetic field.
The excitation power and in-plane magnetic field can therefore be used as control knobs to nucleate vortex cores in ring structures and subsequently generate magnon frequency combs. By optimizing the granularity of the magnetic material and the edge roughness of the structured elements, it should be possible to tune the field range for which the vortex core and, by that, the frequency combs are stable. The design of flat edges which break the rotational symmetry of the vortex can give further control for the direction of static magnetic fields to nucleate a vortex core\cite{dumas2011chirality}.





\section*{Author declarations}

\subsection*{Conflict of Interest}
The authors have no conflicts of interest to disclose.

\subsection*{Author's contributions}

H.S., C.H., and K.S conceptualized the presented work. 
H.S., K.S. and A.K. acquired funding.
K.S. and G.H. fabricated the sample.
A.K., C.H., and J.V.K. performed and analyzed the micromagnetic simulations. 
C.H. carried out the experiments and analyzed the data.
All authors discussed the results.
C.H., H.S., and K.S. visualized the results.
C.H. and K.S. wrote the original draft of the paper. 
All authors reviewed and edited the paper.


\section*{Acknowledgements}

This work received funding from the EU Research and Innovation Programme Horizon Europe under grant agreement no. 101070290 (NIMFEIA) and by the Deutsche Forschungsgemeinschaft (DFG) through the programs GL 1041/1-1 and KA 5069/3-1. Support by the Nanofabrication Facilities Rossendorf (NanoFaRo) at the IBC is gratefully acknowledged.

\section*{Data availability}
The data supporting this study's findings are openly available in RODARE\cite{heins_christopher_2025_3387}. In this work, the authors used scientific colour maps to prevent visual distortion of the data and exclusion of readers with colour-vision deficiencies\cite{crameri_misuse_2020}. Specifically, the used color maps include \textit{oslo}  (https://www.fabiocrameri.ch/colourmaps/) and \textit{cmocean.map} (https://matplotlib.org/cmocean/)).


\bibliographystyle{apsrev4-2}
\bibliography{references.bib}


\end{document}